# Potential Energy Landscape for hot electrons in periodically nanostructured graphene

B. Borca[1], S. Barja[1,2], M. Garnica[1,2], D. Sánchez-Portal[3,4*], V.M. Silkin[3,4,5], E.V. Chulkov[3,4,6], F. Hermanns[1], J.J. Hinarejos[1,7], A.L. Vázquez de Parga[1,2,7*], A. Arnau[3,4,6], P.M. Echenique[3,4,6], R. Miranda[1,2,7]

[1]Dep. de Física de la Materia Condensada, Universidad Autónoma de Madrid, Cantoblanco 28049, Madrid, Spain

[2]Instituto Madrileño de Estudios Avanzados en Nanociencia (IMDEA-Nanociencia), Cantoblanco 28049, Madrid, Spain

[3]Centro de Física de Materiales (CFM-MPC), Centro Mixto CSIC-UPV/EHU, Campus Ibaeta, San Sebastian, Spain

[4]Donostia International Physics Centre (DIPC), Paseo de Manuel Lardizabal 4, 20018 San Sebastian, Spain

[5]IKERBASQUE, Basque Foundation for Science, 48011 Bilbao, Spain

[6]Dep. de Física de Materiales (UPV/EHU), Facultad de Química, Apartado 1072, 20080 San Sebastian, Spain

[7]Instituto "Nicolás Cabrera", Universidad Autónoma de Madrid, Cantoblanco 28049, Madrid, Spain

[*]e-mail: al.vazquezdeparga@uam.es, sqbsapod@sq.ehu.es

ABSTRACT: We explore the spatial variations of the unoccupied electronic states of graphene epitaxially grown on Ru(0001) and observed three unexpected features: the first graphene image state is split in energy, unlike all other image states, the split state does not follow the local work function modulation, and a new interfacial state at +3 eV appears on some areas of the surface. These results show the system behaves as a self-organized periodic array of quantum dots.

Lateral superlattices in graphene, a form of carbon with unique electronic properties,[1] can be realized by *periodic* arrays of doping centres, electronic and geometric corrugations[2] or externally applied potentials[3] and have been predicted to show a number of intriguing properties such as anisotropic propagation of charge carriers,[3] miniband transport and, even, new quasiparticles for triangular superlattices.[4] A lateral superlattice consisting of regions of graphene with different electronic properties can easily be prepared by thermal decomposition of hydrocarbons on single-crystal metallic substrates.[5,6] This process results in an epitaxial graphene monolayer with a periodic array of bumps and valleys (termed ripples) originating from the lattice mismatch between graphene and the different substrates, which provide us with a tuneable pattern of structural and electronic heterogeneities.[7]

Because graphene is only one atom thick it is very sensitive to the interaction with the substrate. Its electronic structure can be modified by the substrate by doping with electrons or holes or by inducing a periodic modulation in the density of states or both effects simultaneously.[8,2] In particular, graphene grown on Ru(0001) is an ideal model system to study the interaction of nanostructured graphene with a metallic substrate because the chemical interaction is spatially modulated with a periodicity of only 29.7 Å.[9,10] Close to the Fermi level this spatially periodic, chemical interaction modulates the electronic structure producing an ordered array of electron pockets.[2] The unoccupied electronic states of these graphene superlattices, however, have not been characterized or discussed so far, in spite of their relevance for the dynamics of excited, hot electrons.

Particularly relevant among the unoccupied states are the image states, bound by the classical image-charge response of metallic surfaces and with a free-electron like dispersion parallel to the surface. The inverse dependence on distance from the surface of the image potential leads to a Rydberg-like series of states that converges to the continuum at the vacuum level ($E_{vac}$).[11] Inverse photoemission studies of the image potential states had shown experimentally that the energy position of the Rydberg series is tied to the local work function of the material.[12,13] In STM experiments the electric field across the tunnel junction causes a Stark shift in the states of the hydrogenic spectrum, expanding the image state spectrum into a resonance spectrum associated with the V-shaped potential. Following the analysis performed by Gundlach in the 60's,[14] the resulting energy spectrum can be approximated by the expression:

$$E_n = \Phi + \alpha(n-0.25)^{2/3} F^{2/3} \quad (1)$$

where, $\Phi$ is the surface work function, $\alpha$ is a constant, F is the electric field between tip and sample and *n* is the quantum number of the states. These Field Emission Resonances (FERs) were experimentally observed in field ion microscopy by Jason[15] and with an STM by Binnig et al.[16] and, since then, they have been used to chemically identify different transition metals at surfaces[17], obtain atomic resolution on insulating surfaces, like diamond[18] or study local changes on the surface work function.[19,20]

In this article we explore by means of scanning tunnelling microscopy/spectroscopy and first-principles calculations the periodic modulation of the unoccupied electronic states in graphene/Ru(0001), mapping the potential energy landscape for hot electrons in this lateral superlattice of graphene dots. Experimentally, we found three unexpected features in this system. The first graphene image state is localized on the H-areas, while it is more extended on the L-areas. This state is *not* tied to the local work function, i.e. it does not shift in energy following the 0.25 eV increase of local work function from L- to H-areas. A new interfacial state at +3 eV appears in the L-areas.

The experiments were done in two independent UHV chambers. One of them equipped with a variable temperature STM working between 80K and 300K, and the other equipped with a Low Temperature STM capable of working at temperatures between 4.6K and 77K. The tunnelling spectra were measured with the feedback loop connected and the variation of the distance between tip and sample, z, was recorded as a function of the bias voltage, V, applied. The z(V) curves were numerically differentiated to obtain the dZ/dV curves shown in Figure 1. Experiments performed using the lock-in technique showed no significant differences in the results.

Figure 1A shows a 150×150 Å$^2$ STM image recorded on a sample of Ru(0001) partially covered with graphene. The image shows the moiré superstructure that appears at this sample voltage as an ordered triangular array of bumps (right half of the image). The local electronic structure, usually determined from the intensity versus voltage characteristics of the tunnel junction, is obtained in this case from the tunnel distance, Z, versus voltage, V, curves, which contain basically the same information. Fig. 1B shows the dZ/dV curve measured on the clean Ru(0001) region. Three peaks reflecting the FERs are visible in the spectra, the energy of the first FER being slightly above the work function (5.4 eV) of Ru(0001).[21] Figure 1C shows the dZ/dV curves measured locally on the H- and L-areas of the graphene moiré (blue and red curves respectively). In the curves measured on the H-areas of the moiré (blue line), where the distance between graphene and the metal substrate is larger, the first FER appears around +4.4 eV, close to the average work function (4.5 eV) measured on this surface.[21]

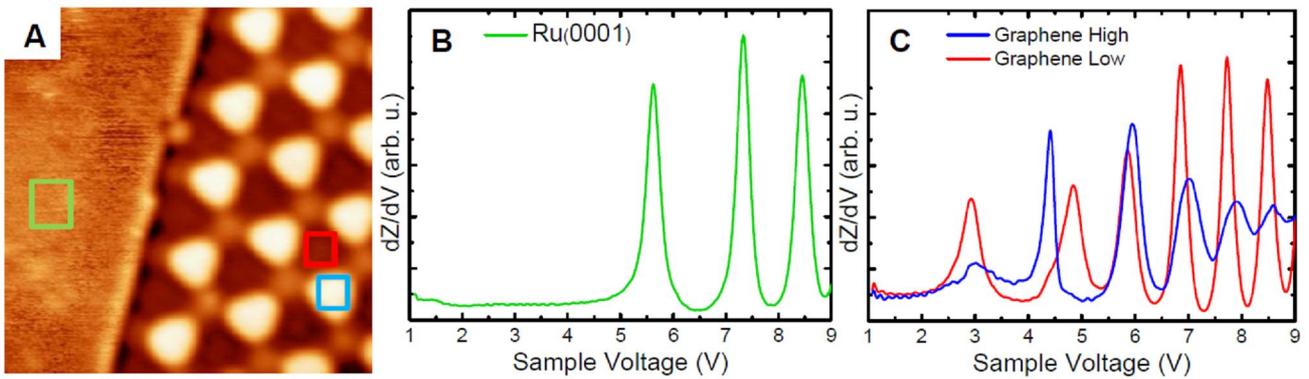

**FIGURE 1**. (A) Scanning tunnelling microscopy image measured at 4.6K at the edge of a graphene island (150×150 Å$^2$, $V_s$= -0.5V). The left hand side is the uncovered Ru(0001) surface, while on the right hand side the moiré superlattice of graphene appears as an triangular array of bumps separated by 29.7 Å. (B) dZ/dV curve measured on the clean Ru(0001) (green box on panel A). The curve shows the presence of the first three FERs. The first one appears at +5.6 eV, slightly above the value of the work function of clean Ru(0001). (C) dZ/dV curves measured on the moiré superstructure on the graphene island. The blue curve is measured in the H-areas of the moiré superstructure (blue box in panel A), it shows the presence of the first four FERs. The red curve shows the corresponding spectra measured in the L-areas of the moiré superstructure (red box in panel A). There are two striking features in this curve, the appearance of a new peak at +3 eV and the shift in energy of the first FER with respect to the one in the H-areas (blue curve) which is in the direction opposite to the change in the local value of the work function between H- and L-areas.

The curves measured on the L-areas of the moiré, where the chemical interaction between graphene and ruthenium is stronger and, accordingly, the interlayer distance shorter, present some unexpected features. First, there is a new peak, 3 eV above the Fermi level, well below the vacuum level, which is not present in the H-areas of the moiré. Second, the first FER (next peak in the curve) is located now at +4.8 eV, i.e. at *higher* energy than in the H-areas. All higher order FERs appear at different energies on L and H-areas, their energy position on L-areas being always smaller than on H-areas, reflecting the difference in the local work function (0.25 eV lower on the L-areas),[22] as expected from theory [see Eq.(1)] and experiments in other systems.[16,18] The first FER at the L-areas shows, however, an energy shift of about 0.4 eV with respect to the peak at H-areas and in the opposite direction to the local work function change. This reversed shift of the first FER is a robust result, which does not depend on the tip sharpness, sample temperature (in the range 4.6K-300K) or tunnelling current used in the experiments, although the exact magnitude of the energy shift in each case depends on the electric field between tip and sample that, in turn, depends on the tip sharpness and the tunnelling current used in the measurements (see Supporting Information).

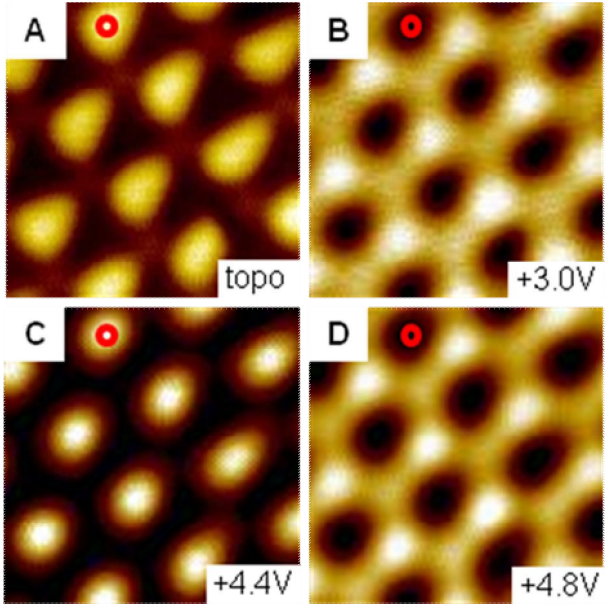

**FIGURE 2.** (A) STM topographic image (90×90 Å$^2$, $V_s$=-0.5V) measured simultaneously with the dZ/dV maps shown in the other panels. (B) dZ/dV map showing the spatial distribution of the new feature that appears at +3V in the L-areas of the moiré superstructure. This feature presents a small modulation in intensity. (C) dZ/dV map showing the spatial distribution of the feature that appears in the spectra a +4.4 V. The intensity of this feature is strictly confined to the H-areas of the moiré. (D) dZ/dV map showing the spatial distribution of the feature that appears in the spectra a +4.8 V, the intensity of this feature is strictly confined to the L-areas of the moiré and presents a small intensity modulation that depends on the stacking sequence of the L-areas of the moiré, similar to that shown in B. The red circle marks the same spot in all the images.

Taking advantage of the spatial resolution of the STM, we map the spatial distribution of the spectral features discussed above. Figure 2A shows a 90×90 Å$^2$ STM topographical image of a single layer of graphene grown on Ru(0001) with the characteristic bright (H-areas) and dark (L-areas) regions. Panels 2B, 2C and 2D show dZ/dV maps measured simultaneously with the STM topography at bias voltages corresponding, respectively, to the peak at +3.0 V and the first FER measured on the H- and L-areas of the moiré at +4.4 and +4.8 V. The dZ/dV map of the new peak at +3.0 V presents intensity only on the L-areas of the moiré. It shows a small modulation, also seen in our topographical images (see Fig. 1 and Fig. 2A), probably due to the different stacking sequence in the L-areas of the moiré,[23] as also observed, for instance, for bilayer Co islands grown on Cu(111) with different stacking.[24] The spatial distribution of the +4.4 eV peak (Figure 2C) reveals that the state is strictly localized on the H-areas. On the contrary, the dZ/dV map at +4.8 eV (Figure 2D) shows localization of the state on the L-areas with a spatial modulation in the amplitude similar to Figure 2B.

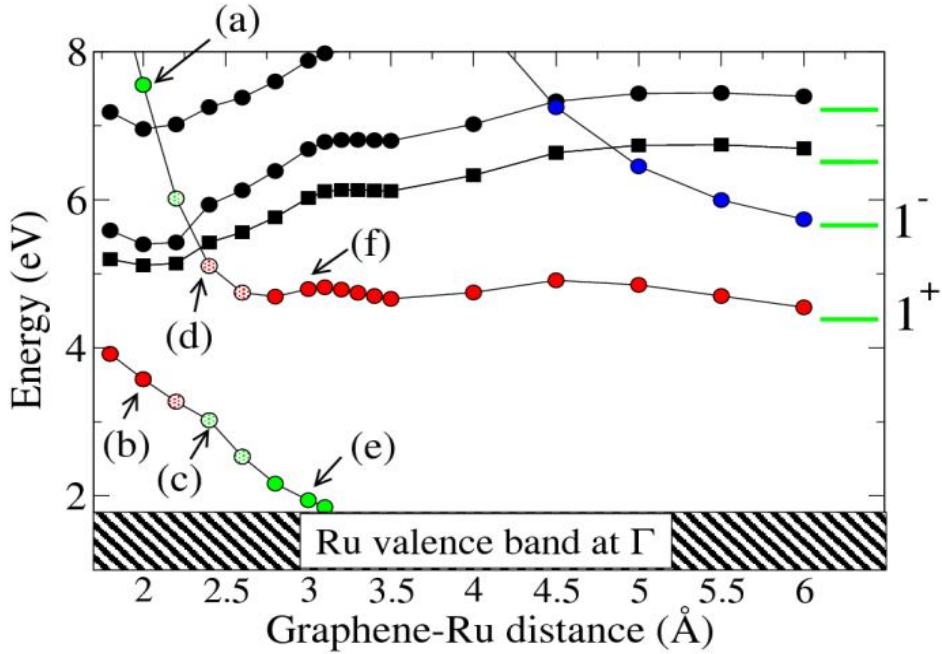

**FIGURE 3.** The figure shows the calculated evolution of the energies (relative to the Fermi level) of the unoccupied states at Γ, i.e., within the projected band gap of Ru(0001), when a laterally strained graphene layer is approached to the surface. The data correspond to the top-hcp stacking of graphene over the Ru surface, but the results are quite independent of the registry between graphene and Ru. The green horizontal lines, at the right hand side, show the energy position of the different states for a free-standing strained graphene layer with our basis set of numerical orbitals supplemented with diffuse orbitals. States labelled with "+1" and "-1" correspond to the first and second image states as described in Ref. (27) The letters refer to the different panels of Figure 4 showing the density associated with particular states. Squares and circles stand, respectively, for doubly and singly degenerate states.

In order to understand the origin of the new spectral feature at +3 eV and the anomalous energy shift of the first FER, as well as the above mentioned different spatial localization of the FERs, we have performed two different sets of first-principles calculations based on density functional theory (DFT). In the first series we have used the SIESTA code[25] to make an explicit description of the electronic and atomic structure of the substrate in order to characterize the (different) C-Ru interactions in the L- and H-areas. A second set of calculations have been performed using plane-waves and pseudopotentials to study the dependence of the field emission resonances on the applied electric field. The metallic substrate has been modelled in this case using a repulsive potential to simulate the effect of the projected band gap around $\overline{\Gamma}$ in the confinement of the graphene image states and field resonances. The first set of calculations permits an unambiguous identification of the resonance appearing at +3 eV on the L-areas, while the second helps in understanding the different way the applied field influences the first FER.

The actual structure of the graphene layer on Ru(0001) involves a very large (and complex) unit cell not yet fully understood.[26] This makes precise first-principles calculations on this system extremely demanding. For this reason, commensurate structures with graphene strained to match the Ru lattice parameter (2.7 Å) were used to identify the main dependence of the electronic structure on the graphene-Ru interlayer distance and registry (see Supporting Information for details). Figure 3 shows the results obtained when a strained graphene plane with a top-hcp registry with respect to the Ru surface layer is moved towards the Ru(0001) surface. Very similar results are found for other stackings and here we just concentrate on the effect of the graphene-Ru distance. First, we will discuss the origin of the different bands. The green horizontal lines in Figure 3 indicate the energy position of the states for a free-standing, but strained, graphene layer. The first two states correspond to the first two "image states" ($1^+$, $1^-$) described in Ref. (27). These states are still bound to the layer using local or semilocal DFT, i.e., without the explicit inclusion of the image-potential tail. At large distances the main effect of the Ru substrate is to break the up-down symmetry of the layer, as well as to confine these extended states due to the gap that exists in the projected band structure at the relevant energy range. As a result, the $1^+$ and $1^-$ states of graphene form linear combinations either with larger weight towards the vacuum (red dots in Figure 3) or in the interface region (blue dots). The later rapidly shift upwards in energy out of the window relevant for our experiments. On the contrary, the $1^+$ state (red dots) stays constant in energy for graphene-Ru distances larger than 3.0 Å.

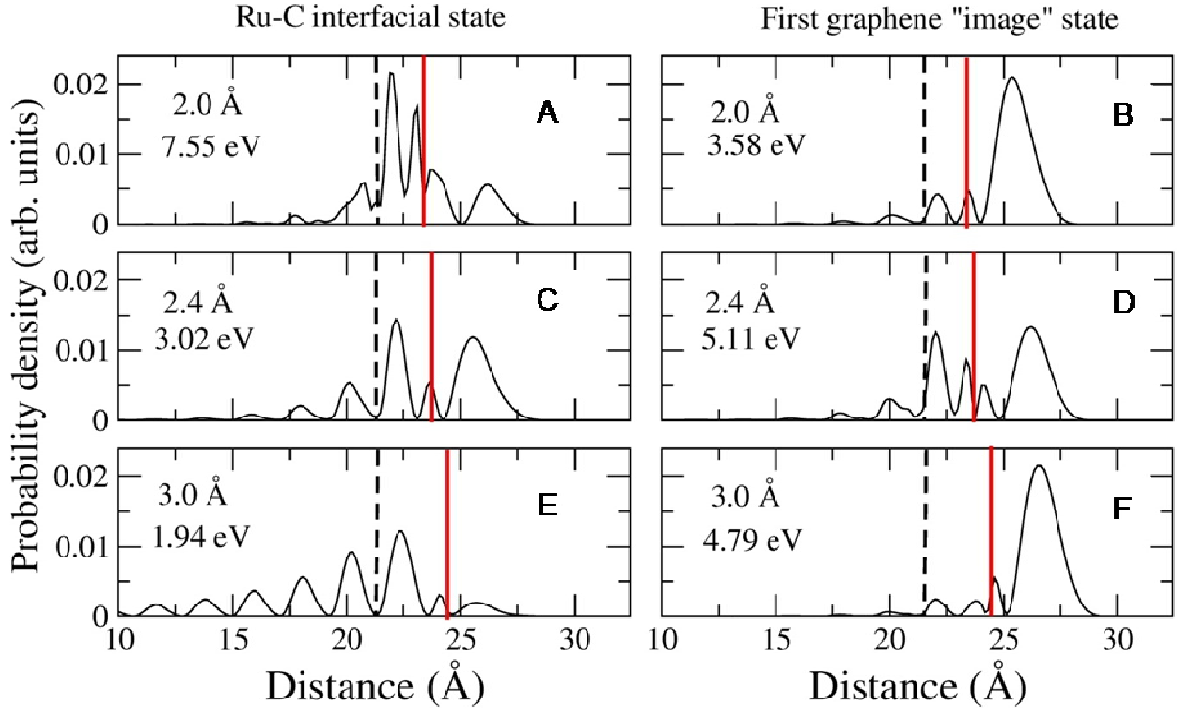

**FIGURE 4.** Planar average of the density associated with selected states in our DFT calculations of strained graphene on Ru(0001). The vertical lines mark the Ru(0001) surface (dashed black) and graphene plane (continuous red). The letters of the panels correspond to the letters shown in Figure 3. For a graphene-Ru distance d ~3 Å (panels E and F), i.e., for relatively weak C-Ru interaction (H-areas), there is no interfacial state, in agreement with STS, while the observed first FER corresponds to the Stark-shifted first image state of graphene ("+1" image state) modified due to the localization induced by the neighbouring Ru surface. For d=3 Å we start to see a Ru(0001) surface resonance emerging from the bottom of the projected band gap. At d=2.4 Å (L-areas), when the interaction is strong (panels C and D), the interfacial state shifts upwards by 1 eV and displaces its charge appreciably outside the graphene layer due to the interlayer confinement, while the first FER shifts upwards by 0.3 eV only and its charge distributes on both sides of the C-surface (both features observed in STS). At d=2.4 Å, both charge distributions at + 3 and +5 eV are quite similar, in agreement with the results shown in Figs. 2B and 2D. A further extra pushing of the C-layer towards the metal to d=2 Å (panels A and B) reveals an avoided

crossing of the two bands and, thus, confirms the origin of the C and Ru states responsible for this interaction.

Approaching the graphene layer further we also induce the confinement of the electronic states of the Ru(0001) surface. This effect is particularly noticeable for a surface resonance of Ru that is pushed to higher energies (green dots in Figure 3). This resonance is reminiscent of the surface states often observed in the (111) surfaces of the noble metals and, for the clean Ru surface, appears slightly below the edge of the Ru(0001) projected band gap. When the distance between the graphene layer and the Ru(0001) surface is smaller than 3.0 Å, this Ru resonance is promoted to a surface state and starts to hybridize with the first image state of graphene (points marked as E and F in Figure 3). The corresponding charge distributions are shown in Figure 4E and 4F. For distances around 2.4 Å this additional resonance, strongly hybridized with graphene, appears near +3 eV (point C in Figure 3). Its corresponding charge distribution is shown in Figure 4C and shows a remarkable increase in the charge above the graphene overlayer. Simultaneously, the $1^+$ image state strongly shifts upwards in energy, as seen in Figure 3 (point D). The corresponding charge distribution is shown in Figure 4D and clearly shows an increase in the charge density in the space between the graphene overlayer and the Ru(0001) surface.

Experimentally we observed a resonance at + 3 eV only in the L-areas. We can now identify it with the rather broad Ru(0001) surface resonance that is lifted upwards (~1 eV from the bottom of the projected band gap at $\overline{\Gamma}$) due to the C-Ru interaction in the strongly bonded L-areas. In general, the adsorption of an adlayer on a metal surface gives rise to the appearance at the interface of new resonances with hybrid character, very sensitive to the strength of the interaction between them.[28] Indeed, recent calculations[29] have shown that the graphene-Ru distance in the L-areas is ~2.4 Å in excellent agreement with our results. Furthermore, similar dZ/dV spectra recorded on the moiré superstructure of graphene on Ir(111), where the graphene-Ir distance is larger, do not show a peak at +3 eV, thus, supporting our identification as an interface peak related to the short C-Ru distance. The corresponding charge distribution is very sensitive to the graphene-Ru distance, as shown in Figure 4. It corresponds to an interfacial state whose charge protrudes outside the graphene layer (allowing its detection by STM) only at short graphene-Ru distances. Below 2.4 Å both the interfacial state (Figure 4A) and the $1^+$ graphene state (Figure 4B) shift to energies higher, particularly the interfacial one than those where they are experimentally detected.

The calculations not only permit the identification of the peak at +3 eV, but also to find the reason behind the reversed shift of the first FER. As the graphene layer approaches the metal surface, the confinement and the hybridization with Ru states shift the $1^+$ state upwards (by ~0.32 eV upon changing the C-Ru distance from 3.0 to 2.4 Å). The applied electrical field can further enlarge the value of the upward energy shift of the first FER in the L-areas (see Supporting Information), where the $1^+$ state appears at higher energies and the local work function is smaller. The combination of these two facts makes the $1^+$ more extended towards the vacuum in the L-areas than in the H- areas and, therefore, more sensitive to the electrical field applied between the STM tip and the surface. This produces an additional upward shift of the first FER in L-areas with respect to H-areas. This relative shift amounts to ~0.5 eV for an electric field of 0.4 eV/Å. Thus, the combined effect of the electric field and the Ru induced confinement over-compensates the change of the local work-function from L to H areas and explains the reversed shift of the first FER.

In summary, the effective potential felt by an electron in the $1^+$ state is less repulsive in the H-areas and can be pictured as a periodic array of shallow attractive (respect to the average) potential wells of nanometer size associated with the H-areas. This explains the strong localization of the peak at +4.4 eV in the H-areas. It is well known that in two dimensions any attractive potential of finite intensity, under quite general conditions, has at least one bound state[30]. The situation for a periodic potential is slightly different,[31] but the effect is similar: the splitting of the bottom of the free-electron-like $1^+$ band into a quantum dot-like lower band (+4.4 eV) localized in the H-areas and a higher band (+4.8 eV) localized in the spatially connected L-areas (see Figure 2).

In conclusion, we have explored the potential energy landscape for hot electrons in a lateral superlattice of graphene grown on Ru(0001). Due to the spatial modulation of the interaction with the metallic substrate, the energy position of the first FER is not tied to the value of the local work function but splits into two sub-bands: the one at higher energy is localized over the extended L-areas and the one at lower energy is localized at the H-areas. The appearance of a new interfacial state was also discussed. The energy position of this state is strongly dependent on the graphene-Ru interaction and, thus, it can be used to calibrate the minimum height of the layer above the Ru surface. The spatial modulation of the first image state of graphene on Ru(0001) resembles that of a periodic array of quantum dots with small binding energies. Therefore, interesting electron correlation effects can be expected when several electrons are simultaneously injected into this band. Selective adsorption of electron donor or acceptor molecules can take place in these superlattices, whose periodicities and potential variations can be tuned by choosing the appropriate substrates. These results could also be important to understand the properties of graphene intercalated compounds[32] where the so called interlayer state, recently identified as the first image state[27], plays a crucial role in the superconducting properties of graphite intercalated materials.[33]

**Acknowledgements**

Financial support by the Ministerio de Ciencia e Innovación through projects CONSOLIDER-INGENIO 2010 on Molecular Nanoscience and grants No. FIS2007-61114, FIS2007-6671-C02-00, and MAT2007-62732, Comunidad de Madrid through the program NANOBIOMAGNET S-2009//MAT1726, the Basque Departamento de Educación, UPV/EHU (Grant No. IT-366-07), the ETORTEK program funded by the Basque Departamento de Industria and the Diputación Foral de Guipuzcoa is gratefully acknowledged.

Supporting Information

# Potential Energy Landscape for hot electrons in periodically nanostructured graphene

B. Borca, S. Barja, M. Garnica, D. Sánchez-Portal, V.M. Silkin, E.V. Chulkov, F. Hermanns, J.J. Hinarejos, A.L. Vázquez de Parga, A. Arnau, P.M. Echenique and R. Miranda

**Influence of the electric field between tip and sample**

The main features present on the dZ/dV curves, i.e. the appearance of a new peak at +3eV in the L-areas of the moiré superstructure and the anomalous (comparing with the high order FERs) shift of the first FER when moving from the H-areas to the L-areas in the moiré unit cell, do not depend on the tip structure, sample temperature or electric field between tip and sample. First, we will focus our attention on the influence of the electric field strength between tip and sample on the dZ/dV curves. In Figure S1 we show two dZ/dV curves measured with two different tunnelling currents using the same tip in the same area of the sample. To exclude tip changes during the measurements, the curves were acquired several times alternating the value of the tunnelling current from one curve to the next. The upper panel in Figure S1 shows the dZ/dV curves acquired with 0.1 nA with tip and sample at 300K. The first FER appears at +4.27 eV in the curves measured in the H-areas of the moiré superstructure and at +4.68 eV in the curves measured in the L-areas. Increasing the tunnelling current by a factor of 30 we are reducing the distance between the tip and the sample and therefore increasing the electric field between them. In the lower panel in Figure S1 we show the curves measured with the same tip in the same sample spot with a tunnelling current of 3.0 nA. In these measurements with a higher electric field between tip and sample, the surface resonance present in the L-areas does not shift in energy and remains at +3.0 eV. The first FER in the H-areas appears slightly shifted in energy (+4.4 eV versus +4.27 eV with low electric field). For the curve measured in the L-areas the first FER appears also higher in energy (+4.9 eV versus +4.68 eV). Thus the anomalous shift of the first FER is a robust feature independent of the strength of the electric field. For the high order FERs the energy shifts are larger, but the relative ordering of the FERs in different regions, with the resonances in the L-areas having always lower energies, is consistent with the observed change of local work function. Notice, however, that this change in behaviour is progressive to a certain extent and the second FER almost does not move when moving from L- to H-areas. Although at higher energies, the second FER still exhibits a non-negligible influence from the potential energy landscape imposed by the graphene-Ru interactions that partially compensates the local work function difference.

It is also interesting to note that there is no appreciable shift in energy of the peak appearing around +3 eV in the L-areas when the electric field between tips and sample is changed, as can be seen in Figure S1. This reinforces our interpretation of this feature as a surface derived state.

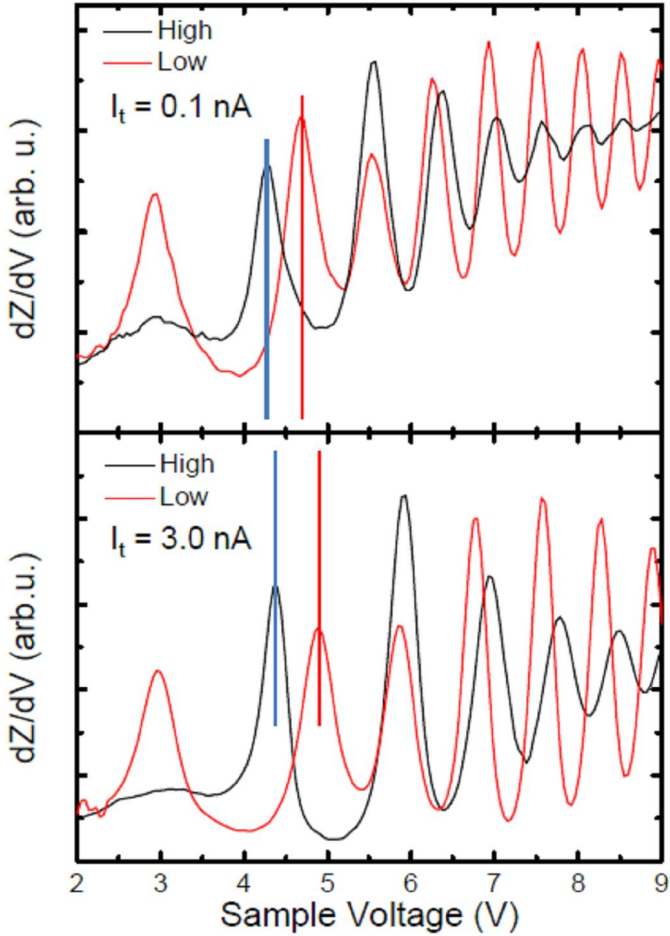

**FIGURE S1.:** dZ/dV curves measured at 300K with the same tip on the same surface spot. The upper (lower) panel shows the curves recorded with a tunnelling current of 0.1 (3.0) nA. The black and red curves reflect the measurements carried out on the H- and L-areas of the moiré superstructure respectively.

**Calculations on strained graphene**

We have performed two different sets of first-principles calculations based on density functional theory (DFT) to compare with our experimental results. In the first series we have used the SIESTA code that uses a basis set of numerical atomic orbitals to represent the electronic states and norm-conserving pseudopotentials to model the interaction of valence electrons with the atomic cores. The first two image states, $1^+$ and $1^-$ according to our nomenclature, are still bound to the graphene layer at the level of standard DFT calculations. However, in order to be able to represent these two states with SIESTA is necessary to include diffuse orbitals with s and p symmetry to the Carbon basis set. The purpose of the SIESTA calculations is to explore the dependence of the interaction between the Ru surface and the graphene layer with their interlayer distance and, in particular, the possible appearance of interface states due to the hybridization of Ru and graphene states. Since the complex relationship with the substrate and the large unit cell of the moiré (periodicity of the superlattice, 29.7 Å) would be very difficult to simulate accurately by first principles calculations[1,2] we have adopted a different strategy. We have performed calculations for a 1x1 cell of strained graphene with a lateral lattice parameter of 2.7 Å and, thus, commensurate with the Ru substrate as a function of the graphene-Ru distance. A similar approach has been followed in some other recent works.[3] We have checked before that the band structure of this strained graphene is qualitatively identical to that of relaxed graphene. Furthermore, the results are quite insensitive to the registry of graphene on the Ru surface layer, but strongly dependent on the graphene-Ru distance. In our calculations we have used a symmetric slab containing 23 layers of Ru with a graphene layer on top of both surfaces.

**Model calculation with plane waves**

A second set of calculations have been performed using plane-waves and pseudopotentials to study the dependence of the Field Emission Resonances on the applied electric field. The metallic substrate has been modelled in this case using a repulsive potential to simulate the effect of the projected band gap around Γ on the confinement of the graphene image states and field resonances. The upper panels in Figure S2 show the energy position and the charge distribution for the $1^+$ state without electric field for two different configurations, the one at the left with the graphene closer to the metal surface (L-area) and the one at the right with the graphene further away (H-area). Changing the distance between graphene and the metallic substrate strongly modifies the charge distribution. For larger distances (right panels) the charge density increases in the area between the graphene and the ruthenium surface. The energy position of the state shifts down when the distance is increased. In the L-areas the $1^+$ state appears at higher energies and the local work function is smaller. The combination of these two facts makes the $1^+$ more extended towards the vacuum in the L-areas than in the H-areas and, therefore, more sensitive to applied field.

Note that this effect cannot be observed in Figure 4 of the main text. The decay into the vacuum of the charge distributions shown there only reflects the shape of the diffuse orbitals used to augment the basis set of local orbital in those calculations. We have checked, however, using our model calculations with plane-waves that the effect due to the applied electric field between the STM tip and the surface produces an additional upward shift of the first FER in L-areas respect to H-areas.

If an electric field is applied to the system there are some major changes in the energy positions and also in the charge distribution (Figure S2 lower panels). When the distance between graphene and ruthenium is small (i.e. the L-areas) the main effect of the electric field is to Stark shift the image state upwards in energy by more than 0.5eV, but the charge distribution is almost identical with and without electric field (left panels). On the contrary, when the distance between graphene and ruthenium is larger the electric field modifies the charge density distribution, increasing it between the graphene layer and the ruthenium and reducing its amplitude towards the vacuum (Figure S2 right panels). In this case there is a shift to higher energies smaller than 0.1 eV. Furthermore, simple comparison of the two lower panels in Figure S2 reveals that, in the presence of applied field, the energy shift from H- to L-areas amounts approximately 0.5 eV in agreement with the observations.

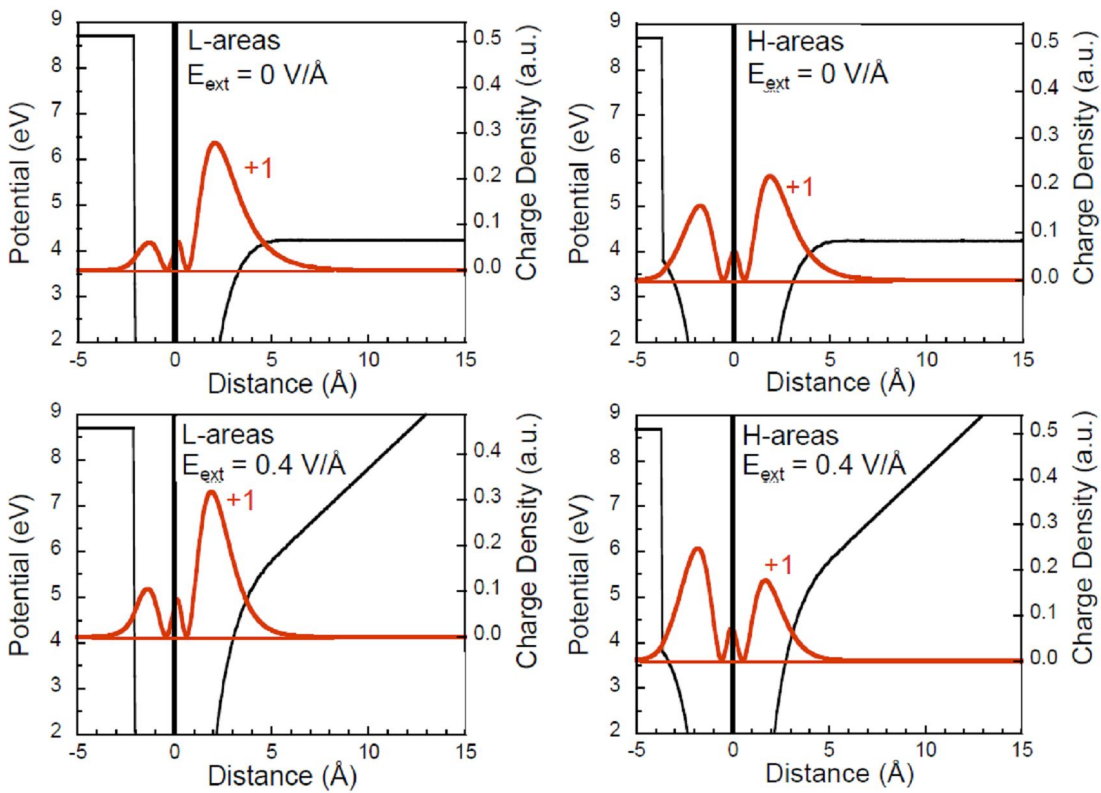

**FIGURE S2:** The upper panels show the energy position (thin red horizontal line) and charge density of the $1^+$ image state without external electric field for relative distances between graphene and ruthenium corresponding to the L and H-areas respectively. The lower panels show the same configuration but with an external electric field of 0.4 V/Å. The graphene layer is placed at distance d= 0 Å. (thick vertical black line).